%% file: elliu.tex
\documentclass[12pt]{article}

\input preamble.tex

\title{Evaluating Deep Learning Models and Adversarial Attacks on 
Accelerometer-Based Gesture Authentication}

\author{Elliu Huang\footnotemark[1]\ \ \  Fabio Di Troia\footnotemark[1]\ \ \ 
Mark Stamp\footnotemark[1]\,\,\footnotemark[2]}

\begin{document}

\symbolfootnotetext[1]{Department of Computer Science, San Jose State University}
\symbolfootnotetext[2]{mark.stamp$@$sjsu.edu}

\maketitle

\abstract
Gesture-based authentication has emerged as a non-intrusive, 
effective means of authenticating users on mobile devices. Typically, such 
authentication techniques have relied on classical machine learning techniques, 
but recently, deep learning techniques have been applied this problem. 
Although prior research has shown that deep learning models are vulnerable to 
adversarial attacks, relatively little research has been done in the adversarial 
domain for behavioral biometrics. In this research, we collect tri-axial accelerometer 
gesture data (TAGD) from~46 users and perform 
classification experiments with both classical machine learning 
and deep learning models. Specifically, we train and test
support vector machines (SVM) and convolutional neural networks (CNN).
We then consider a realistic adversarial attack, 
where we assume the attacker has access to real users’ TAGD data, 
but not the authentication model. We use a deep convolutional generative 
adversarial network (DC-GAN) to create adversarial samples, 
and we show that our deep learning model is surprisingly robust 
to such an attack scenario.

\section{Introduction}\label{chap:introduction}

With the ubiquity of technology, authentication has become an essential part of everyday life. Passwords and PINs are the most common forms of authentication, but biometrics are also popular. Biometric authentication includes physiological (e.g., facial recognition and fingerprint) and behavioral (e.g., gait and keystroke dynamics) approaches~\cite{bhattacharyya2009biometric}. 

While physiological biometric authentication has proven to be highly effective, sensors and equipment required for such approaches are usually costly. Additionally, attackers can sometimes bypass such a system if they have access to a copy of the required features~\cite{wu2020liveness}. On the other hand, behavioral biometrics not only have the potential to be cost effective, they may also be more secure, at least in cases where attackers have difficulty imitating the relevant features. Furthermore, the non-intrusive nature of behavioral biometrics may be considered desirable, in comparison to physiological biometric authentication.

Gesture-based authentication is a relatively recent behavioral biometric that has achieved promising results. There are various techniques for analyzing gestures, including acceleration, angular motion, 3D motion, and a mix of the three. Several machine learning techniques, including those we discuss in Section~\ref{chap:background}, have been applied to the gesture-based authentication problem. 

In this research, we explore the effectiveness of deep learning techniques on gesture-based authentication. Our research is based on a new dataset that we have collected. 
Given that our tri-axial accelerometer gesture data (TAGD) are time series, we consider two time series classification (TSC) techniques: support vector machines (SVM) and one-dimensional convolutional neural networks (1D-CNN). When combined with feature extraction techniques, our SVM model provides for rudimentary analysis of our TAGD data, as well as a basis for comparison to our 1D-CNN model. We also generate adversarial samples using generative adversarial networks (GAN) and use these samples to explore the robustness of our 1D-CNN model against a realistic adversarial attack.

The remainder of this paper is structured as follows.
Section~\ref{chap:background} discusses relevant work
in the field of gesture-based authentication and adversarial attacks on biometric authentication. 
Section~\ref{chap:dataset} provides an overview 
of the dataset, including specific steps of the data collection process and data preprocessing techniques. 
Machine learning techniques, including generating adversarial samples, 
and adversarial strategies are introduced in Section~\ref{chap:implementation}. 
We present our experimental results for our classification models and an adversarial attack in Section~\ref{chap:results}. Our conclusion and a brief discussion of future research directions
are provided in Section~\ref{chap:conclusion}.

\section{Related Work}\label{chap:background}

Relative to the vast research literature on behavioral biometrics,
there is comparatively little work on gesture-based authentication.
In this section, we provide an overview of research on gesture-based
authentication and adversarial attacks on such security systems.

Most gesture-based authentication techniques can be categorized into two main methods,
namely, touchscreen and motion gestures~\cite{clark2015engineering}. There are, however, 
studies that combine both into a single authentication system~\cite{buriro2016hold}. 

Touchscreen-based gesture authentication methods typically analyze touch dynamics, 
including various inputs recorded from a touchscreen interface such as finger size and pressure. 
One approach consisted of collecting finger behavior and position data and authenticated users via SVMs~\cite{alariki2014touch}. Another study employed  
particle swarm optimization to find patterns in touchscreen dynamics~\cite{meng2012touch}.

Motion gestures generally rely on accelerometer and gyroscope data to analyze 
the acceleration and angular motion of the mobile device. Prior research in this
domain has applied dynamic time warping (DTW)~\cite{liu2009uwave}, 
SVMs~\cite{lu2017data}, and hidden Markov models (HMM)~\cite{guse2017gesture} 
to authenticate users. One gesture-based approach employed a more sophisticated 
method that involved the ``leap motion'' controller that collects 3D motion 
data and applied similarity thresholds to authenticate users~\cite{imura2018hand}. 
Another approach analyzed full-body and hand-gestures 
in 3D space using two-stream CNNs~\cite{wu2016two}.

Adversarial attacks on gesture-based authentication is not a well-researched area, 
but there are a handful of relevant studies in the general field of behavioral biometrics. 
One study analyzed behavioral mouse dynamics and found that deep learning 
authentication models were susceptible to adversarial attacks~\cite{tan2019adversarial}. 
Adversarial samples have been generated using a fast gradient sign method (FGSM) 
to create perturbations in the data, with a gated recurrent unit (GRU) then used 
to generate adversarial samples. Another study~\cite{agrawal2021defending} analyzed the 
resilience of continuous touch-based authentication systems (TCAS) to adversarial attacks. 
These researchers found that their TCAS trained with the help of generative adversarial networks (GAN) 
had a lower false acceptance rate than that of vanilla TCAS.
The paper~\cite{al2016reconstruction} reports on
experiments with randomization attacks on gesture-based security systems that use SVMs,
and finds that their models are highly vulnerable to adversarial attacks. 

Similarly, adversarial learning on time series classification has not seen much research. 
Most adversarial attacks involve small perturbations 
of the original data using state-of-the-art FGSM or basic iterative method (BIM) in order 
to ``trick'' a classification or regression model. A study of 
adversarial attacks on multivariate time series regression found that three of the most popular deep learning 
models---CNNs, GRUs, and long short-term memory (LSTM)---were highly susceptible to such attacks~\cite{mode2020adversarial}. Another study also used FGSM and BIM to create small 
perturbations in time series data, which significantly lowered the classification accuracy
of deep learning models for vehicle sensors and electricity consumption data~\cite{fawaz2019adversarial}.

\section{Dataset}\label{chap:dataset}

In this section, we give an overview of the data collected and specify the
steps in the data collection process. We also include a discussion of data
preprocessing and the feature engineering techniques that we have employed.

\subsection{Data Collection}

\begin{figure}[!htb]
   \centering
   \begin{tabular}{ccc}
    \includegraphics[width=0.45\textwidth]{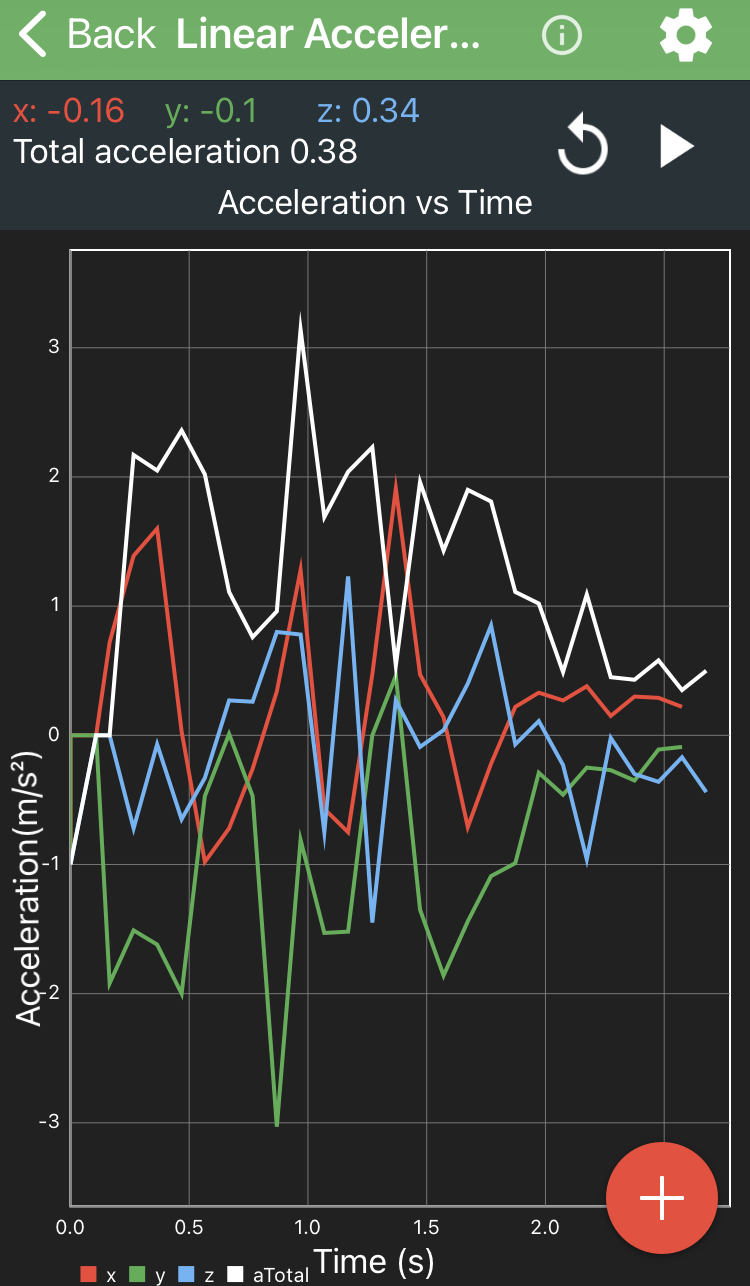}
    &
    &
    \includegraphics[width=0.45\textwidth]{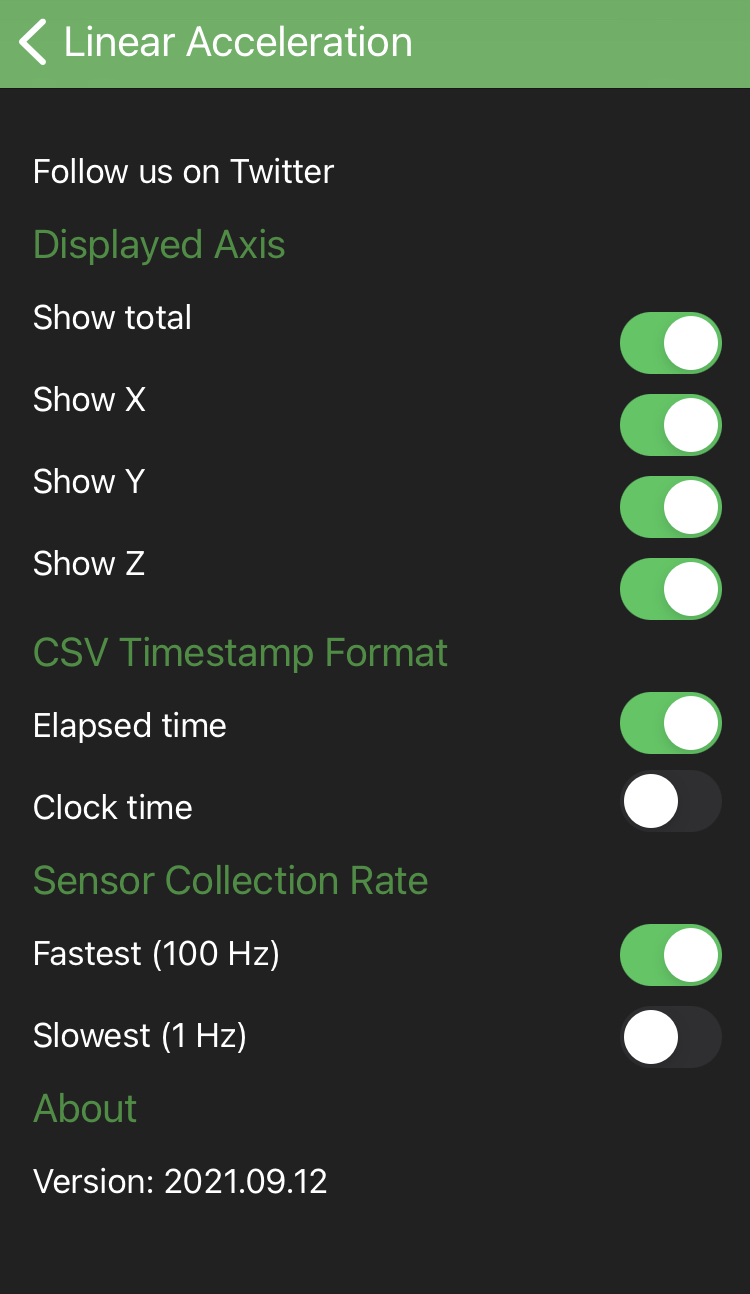}
    \\
    (a) Accelerometer sensor
    &
    &
    (b) App settings
    \end{tabular}
    \caption{Screenshots of Physics Toolbox Sensor Suite app}\label{fig:1}
\end{figure}

In this research, we collect users’ tri-axial accelerometer gesture data (TAGD) 
while the user holds a smartphone and writes their ``signature'' in the air, 
similar to the process in~\cite{huang2021new}. The accelerometer sensor of 
the Physics Toolbox Sensor Suite~\cite{iosapp}
was used to collect data. A screenshot of the application is shown in 
Figure~\ref{fig:1}.
The tri-axial acceleration is represented as separate 
curves---the red curve represents acceleration
along the x-axis, the green curve represents acceleration along the
y-axis, the blue curve represents acceleration along the z-axis, and
the white curve represents the total magnitude of acceleration. Data
is collected at a frequency of~100 Hz, i.e., 100 data points per
second. We performed data collection solely on the Apple~iOS platforms in order
to reduce the possibility of smartphone type being a confounding variable.
We note that iPhone models varied from iPhone 8 to iPhone~X. 

After installing the app,
he user performs the following steps to collect signature data.
\begin{enumerate}
    \item Tap the red button to start recording accelerometer data
    \item Move the smartphone in the air to draw a signature
    \item Tap the red button again to stop recording data
    \item Upload data in the form of a CSV file into Google Drive
\end{enumerate}
These steps were repeated~50 times to collect~50 signatures for each user. 

\begin{figure}[!htb]
    \centering
    \includegraphics[width=0.75\textwidth]{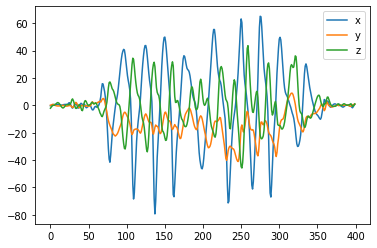}
    \caption{Sample tri-axial acceleration time sequence}
    \label{fig:2}
\end{figure}

In total, we collected~50 signatures from each of~46 different users.
Users typically chose their initials as their signature, but they were free to 
create their own unique signature. Generally, the time to write each signature 
varied between~3 and~7 seconds, and the entire data collection process 
for one individual required about~20 minutes. 
Our dataset is freely available for use by other researchers 
at~\cite{dataset}. A sample of our TAGD data is shown in Figure~\ref{fig:2}.

\subsection{Data Preprocesssing}

The raw tri-axial data is of variable length. As discussed below,
we resize all signatures to the same 
size, as most traditional classification techniques require input data of a fixed size. 
We have applied several feature engineering techniques to the resulting time series.

\subsubsection{Feature Engineering}\label{sect:FE}

As noted above, each signature is a temporal sequence of tri-axial accelerometer data. 
First, we extract statistical features based on the the acceleration for each axis, 
ignoring the sequential nature of the data. The resulting distributions 
vary---we compute the following statistical measures of shape, 
center, and spread for each of the three axes.

\begin{description}
\item[\bf Length ($L$)]\hspace*{-12pt} --- The number of data points in the signature. 

\item[\bf Mean ($\mu$)]\hspace*{-12pt} --- The center of the distribution is the mean
$$
  {\mu}= \frac{1}{n} \sum_{i=1}^{n} x_{i} = \frac{1}{n}(x_1 + x_2 + \cdots + x_i)
$$

\item[\bf Median ($m$)]\hspace*{-12pt} --- The median is another measure of the center of the distribution.

\item[\bf Standard deviation ($\sigma$)]\hspace*{-12pt} --- The standard deviation 
$$
  \sigma = \sqrt{\frac{\displaystyle\sum_{i=1}^{n} (x_{i}-\mu)}{n-1}}
$$
measures the variability in the signature data.

\item[\bf Kurtosis ($k$)]\hspace*{-12pt} --- The kurtosis is computed as
$$
    k = \frac{\displaystyle\sum_{i=1}^{n}(x_{i}-\mu)^4}{n \sigma^4}
$$
and it measures the weight of the tails relative to the center of
the distribution and provides additional information related to the  
signature motion.

\item[\bf Skewness ($s$)]\hspace*{-12pt}--- The symmetry of the distribution, which
is computed as 
$$
    s = \frac{\displaystyle\sum_{i=1}^{n}(x_{i}-\mu)^3}{n \sigma^3}
$$
can help us understand the ``smoothness'' of motion in a signature.
\end{description}
All of these features provide some information 
about underlying patterns in users’ signatures. 

For each signature, we calculate a feature vector of the form
$$
  (L, \mu_{x}, \mu_{y}, \mu_{z}, 
  	m_{x}, m_{y}, m_{z}, 
	\sigma_{x}, \sigma_{y}, \sigma_{z}, 
	k_{x}, k_{y}, k_{z}, 
	s_{x}, s_{y}, s_{z})
$$ 
consisting of the measures of shape, center, and spread of the distribution, 
as discussed above. This feature vector of~16 elements is utilized only in 
our SVM experiments, below.

\subsubsection{Time Series Resampling}\label{chap:tslearn}

Since 1D-CNNs and GANs require feature vectors of fixed length, we 
use \texttt{tslearn}~\cite{tavenard2021tslearn}, we resize all the TAGDs to length~400. 
The resizing function in \texttt{tslearn} interpolates for arrays less than the target size. 
We chose to resample all the time series to 
length~400 since the median 
length of the sequences is~380, 
while the mean length is~400.

\section{Implementation}\label{chap:implementation}

In this section, we discuss the machine learning techniques that form the basis 
of our experiments. 
These techniques, namely, support vector machines, convolutional neural networks, and
generative adversarial networks,
will serve as the basis for our classification experiments in 
Section~\ref{chap:results}. 
We also introduce our strategies for adversarial attacks.

\subsection{Support Vector Machines}

\begin{figure}[!htb]
    \centering
    \input{figures/kernel_trick.tex}
    \caption{The kernel trick}\label{fig:3}
\end{figure}
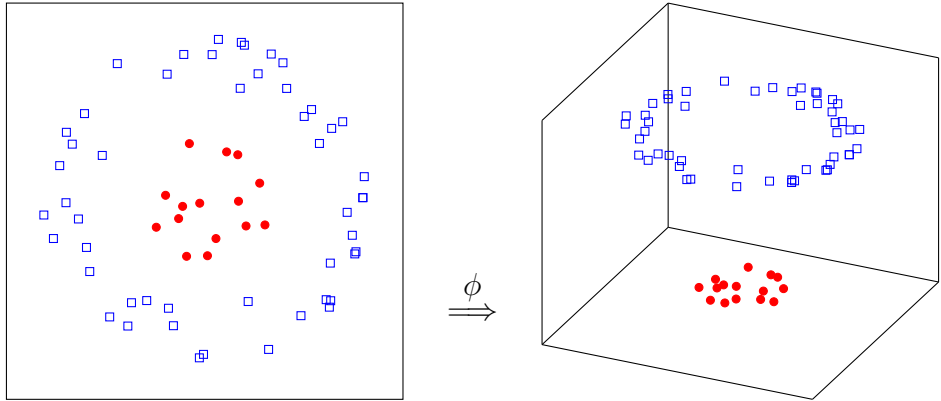

Support vector machines (SVMs) are one of the most popular supervised learning techniques 
for classification and regression. SVMs attempt to find the optimal separating hyperplane between 
two labeled sets of training data~\cite{StampMark2018Itml}. However, a dataset 
need not be linearly separable, in which case we can employ the ``kernel trick.'' As depicted in Figure~\ref{fig:3},  
the kernel trick maps the input data into a higher-dimensional space where it is more likely to 
be linearly separable. The kernel trick, together with ``soft margin'' calculations
that allow for classification errors, makes an SVM an extremely powerful and flexible tool 
in the field of machine learning.

A classification problem with a small training sample size and high dimensionality is prone to overfitting~\cite{chen2007enhanced}. Feature selection techniques can help to
prevent this problem by discarding features, with minimal loss---or even improvements---in performance. 
In our SVM experiments, we used support vector machine recursive feature elimination (SVM-RFE) 
for feature selection. SVM-RFE consists of eliminating the least significant feature (based on
linear SVM weights), then training a model on the reduced feature set. This process is repeated
until the desired number of features is reached, 
or the performance degrades beyond acceptable limits.

\subsection{1D Convolutional Neural Networks}\label{chap:cnn}

Typically, convolutional neural networks (CNN) are associated with feature extraction and 
classification for images, which generally involves two-dimensional convolutional neural networks (2D-CNN).
In such a model, 2D data (e.g., images) are fed into a CNN and classified via a final fully-connected layer. 

In this paper, we do not consider images; instead, we have temporal sequences of fixed length.
Such data is suitable for one-dimensional convolutional neural networks (1D-CNN). 
While not as common as 2D-CNNs, 1D-CNNs are used for signal processing and 
sequence classification, with numerous applications in biomedical and civil engineering~\cite{kiranyaz20211d}.

The architecture of a 1D-CNN is analogous to that of a 2D-CNN, with the key differences being 
the dimensionality of the input data and the convolution operation. 
2D-CNNs typically use a rectangular kernel that slides from left to right, top to bottom. 
In contrast, 1D-CNNs employ a kernel that spans some number of variables and slides along 
a vector.

\subsection{Adversarial Strategy}

We also consider how our deep learning models perform under adversarial attacks. 
While several studies analyze adversarial attacks involving small perturbations of the original data~\cite{ismail2019adversarial}, we explore a scenario where we assume the intruder 
has access to a real users’ gesture data. We test both poisoning and evasion attacks
using learning models to generative adversarial samples; specifically, we use a type of
generative adversarial network (GAN) to produce adversarial samples. 

\subsubsection{Deep Convolutional Generative Adversarial Networks}

Two competing neural networks are trained in a GAN---a generative network and a 
discriminative network---with 
the generative network creating fake data that 
is designed to defeat the discriminative network. The two networks are
trained simultaneously following a game-theoretic approach.
In this way, both networks improve, with the ultimate objective being a
model (discriminative, generative, or both)
that is stronger than it would have been if it was trained only 
on the real training data; see~\cite{creswell2018generative}
for additional details.

\begin{figure}[!htb]
    \centering
    \includegraphics[width=0.9\textwidth]{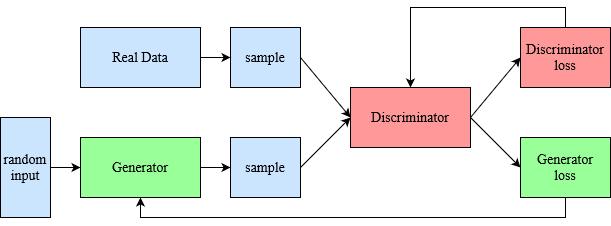}
    \caption{Overview of GAN structure}
    \label{fig:6}
\end{figure}

An overview of GAN structure is depicted in Figure~\ref{fig:6}. 
Among many other uses, GANs have been used to generate realistic 
adversarial samples.

We use deep convolutional GANs (DC-GAN) to replicate our time series data in a form
that will serve as adversarial samples~\cite{barua2019fcc}. The generator 
and discriminator models are both based on 1D-CNN models. 
The generator essentially performs the functions of a 
convolutional layer in reverse---the input is an arbitrary sequence of values and it uses 
transposed convolution layers to shape the data into a desired form. The discriminator can be 
based on a traditional convolutional neural network. The fully-connected layer outputs a 
value between $\hbox{}-1$ and $\hbox{}+1$, with a negative output indicating a fake sample 
and positive output indicating a real sample. The generator and discriminator are connected
by a loss function, which provides feedback to both models. Over several epochs of training, 
the generator should becomes better at generating adversarial samples while the 
discriminator should become better at distinguishing between real and fake samples.

In our GAN model, the generator has an input consisting of a sequence of~100 values 
sampled from a normal distribution. After the data passes through
three transposed convolution layers, 
the sequence of~100 values are transformed into a sequence of the same size 
as the TAGD, that is, $400\times 3$. The discriminator model closely follows the architecture 
of the 1D-CNN classification model outlined above, with the major 
difference being the binary output of the fully-connected layer. The generator and discriminator 
are connected by a binary cross entropy loss function. In our experiments, we 
vary the number of training epochs to see how effective the adversarial samples 
are in breaking down our model.

\begin{figure}[!htb]
   \centering
   \begin{tabular}{ccc}
    \includegraphics[width=0.45\textwidth]{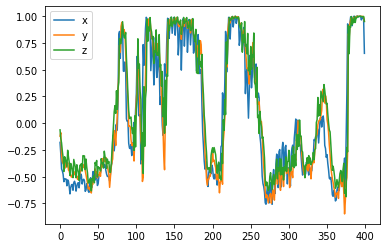}
    &
    \includegraphics[width=0.45\textwidth]{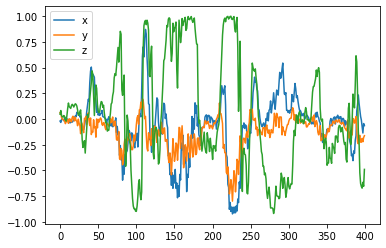}
    \\
    (a) Acceleration sequence after~10 epochs
    &
    (b) Acceleration sequence after~25 epochs
    \\
    \includegraphics[width=0.45\textwidth]{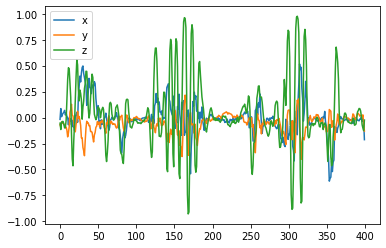}
    &
    \includegraphics[width=0.45\textwidth]{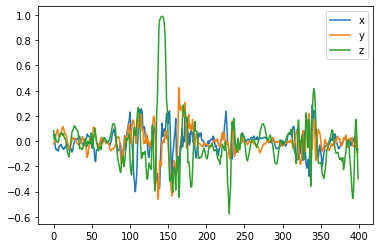}
    \\
    (c) Acceleration sequence after~50 epochs
    &
    (d) Acceleration sequence after~100 epochs
    \end{tabular}
    \caption{DC-GAN generated acceleration sequences}\label{fig:10}
\end{figure}

In Figure~\ref{fig:10}, we have examples of TAGD generated using DC-GANs with different training epochs. The number of training epochs is directly related to how well the the DC-GAN model can replicate data. As we can see in Figure~\ref{fig:10}~(a) and (b), the data is quite random doesn't resemble the real TAGD in~\ref{fig:2}, whereas Figures~\ref{fig:10}~(c) and (d) appear closer to the real TAGD. As we train the DC-GAN more epochs, the adversarial samples resemble the real data more. 

\subsubsection{Adversarial Attack}\label{chap:poison}

Similar to~\cite{munoz2019poisoning}, we ``poison'' our training dataset with adversarial samples 
generated from DC-GANs, meaning that we mix in adversarial samples with our real training dataset. 
Then, we train our 1D-CNN on the poisoned dataset and try to classify real data. The accuracy of 
classifying with the poisoned training dataset suggests how well our 1D-CNN can survive a poisoning 
attack while simultaneously indicating how well our DC-GAN can generate adversarial samples.

\section{Experiments and Results}\label{chap:results}

In this section, we present and analyze the results of the experiments outlined in the previous section. 
For our first experiment, we provide the results of a multiclass classification problem using SVMs, 
based on the statistical features discussed in Section~\ref{sect:FE}. 
Then we apply a deep learning technique, 1D-CNNs, 
in the multiclass classification problem. Both of these techniques, SVMs and 1D-CNNs, 
produce strong results. Then we move on to adversarial learning where we use GANs to generate 
adversarial samples. With the GAN-generated adversarial samples, we show that our deep learning 
model is robust under a poisoning type of adversarial attack. 

We use several metrics in our multiclass classification problem. The most basic metric we 
consider is the accuracy, which is calculated from a~46 by~46 confusion matrix. 
Of course, higher accuracy indicates a more successful model. We also measure the
false acceptance rate (FAR), which is the rate at which a different user is classified as the actual user, 
and the false reject rate (FRR), which is the rate at which real users are mis-classified as other users. Since we are dealing with multiclass classification, we calculate FAR and FRR for each individual user and report the average for the 46 different users~\cite{grandini2020metrics}. 
Intuitively, lower FAR and FRR indicates greater success in the classification model. 

In our SVM experiments, we use SVM and RFE libraries from \texttt{scikit-learn}. 
In our 1D-CNN and GAN experiments, we implemented \texttt{Keras} libraries~\cite{brownlee2016deep} 
to develop our models. For all the experiments, we use an~80-20 train-test split,
i.e. 80\%\ of the data is used to train the model, while the remaining 20\%\ is used for testing.

\subsection{SVM Results}

We first considered rudimentary experiments with various kernels and values of parameters 
and found that a linear kernel with regularization parameter~$C=1000$ 
worked the best for multiclass classification. 
As a result, for all experiments in this section, we use an SVM with linear kernel and~$C=1000$. 

Here, we train our SVM model on the feature vector described in 
Section~\ref{sect:FE} and use SVM-RFE to select the strongest features. We
analyze the relationship between the number of features and the FAR, FPR, and accuracy.

\begin{figure}[!htb]
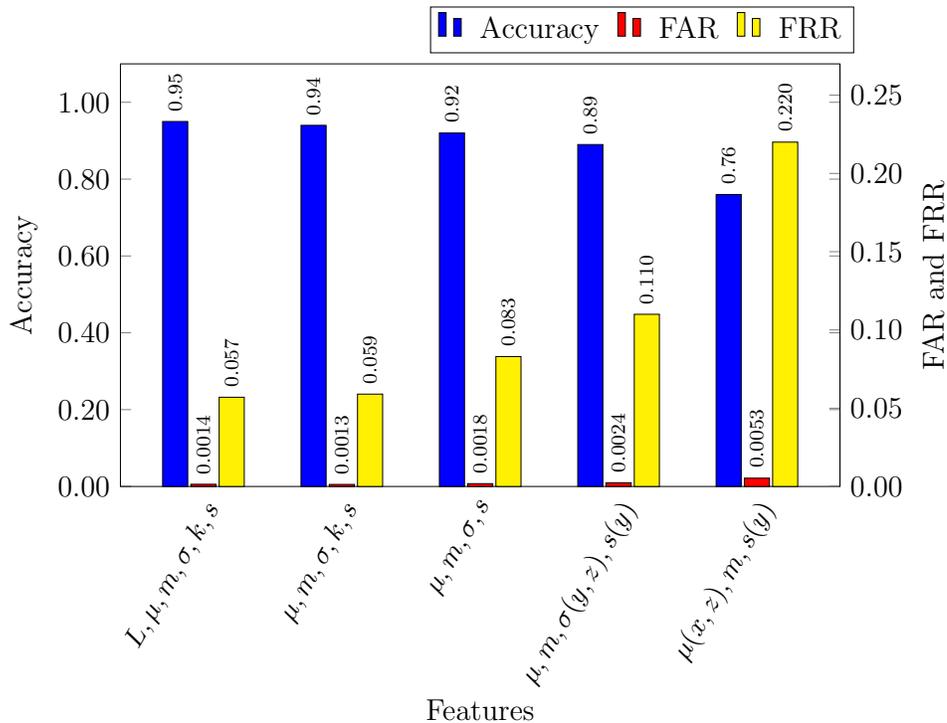

\input figures/barMulticlassAccuracy.tex
\caption{Linear SVM results for selected combinations of features}\label{fig:7}
\end{figure}

Using all~16 features discussed in Section~\ref{sect:FE}, 
the SVM achieved a~95\%\ classification accuracy and~0.0014 and~0.057 FAR and FRR, 
respectively. These results (and more) are summarized in Figure~\ref{fig:7}. 

As we eliminate features based on the rankings determined by SVM-RFE, the classification 
accuracy in Figure~\ref{fig:7}~(c) generally decreases, dropping to about~89\%\ 
with~9 of the~16 features. This is still quite strong, considering the number of classes. 
As we can see from Figure~\ref{fig:7}, the accuracy 
generally decreases slightly as we eliminate more features, although the accuracy does not 
drop below~90\% until we have eliminated~7 features. Similarly, the FAR is exceptionally 
low even as features are eliminated, staying below~1\%\ for every combination of features. 
However, the FRR starts at around~5\%\ and increases to more than 20\%\ once
we have eliminated~10 features.

\subsection{1D-CNN Results}\label{chap:cnnresults}

\begin{figure}[!htb]
    \centering
    \includegraphics[width=0.9\textwidth]{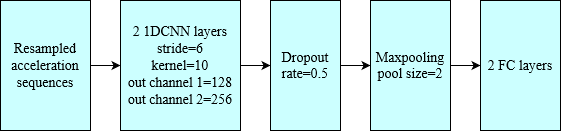}
    \caption{1D-CNN architecture}\label{fig:1dArch}
    \label{fig:8}
\end{figure}

We also experiment with a 1D-CNN as our classification model,
where the model follows the architecture in~\cite{brownlee_2020}. 
The 1D-CNN model is trained on temporal sequences of fixed length, as described in Section~\ref{chap:implementation}. The model performs one-dimensional convolutions along 
the time axis with RELU activation functions after each convolution layer. 
The first convolution layer produces~128 filters, while the second convolution 
layer produces~256 filters. A dropout layer follows the convolution layers to prevent overfitting~\cite{hinton2012improving}. Then we have a~1D max-pooling layer to 
downsample the data and highlight the key features. 
The output of the max-pooling layer is passed through a flatten layer, 
which is then passed through two fully-connected layers. 
The last fully-connected layer produces the value that corresponds to the classification.
This model is illustrated in Figure~\ref{fig:1dArch}.

After fine-tuning hyper-parameters (dropout rate and number of filters), 
we moved on to experiment with kernel size and stride length. We found
the accuracy hovered around~89\%\ to~91\%. According to~\cite{tang2020rethinking},
the performance of our 1D-CNN should be more receptive to changes in hyper-parameters 
involving the convolution layers.

We tested different combinations of kernels and stride lengths in the convolution layers. 
These results are summarized in Table~\ref{tab:stride}. We see that most of the results are 
fairly strong, ranging from a low accuracy of~90\%\ to a high of~94\%. 
Generally, as the kernel size increases, the accuracy increases. 
Similarly, as the stride length increases, the accuracy tends to increase. 
This is most likely due to the fact that larger kernels and stride lengths produce more refined 
features for the fully-connected layers. 

\begin{table}[!htb]
\advance\tabcolsep by 2pt
\caption{Stride length versus kernel size}\label{tab:stride}
\centering
\begin{tabular}{cclll}
\midrule\midrule
  \multirow{2}{*}{stride\ \ }
  & \multicolumn{4}{c}{kernel} \\ \cmidrule{2-5} 
  & \multicolumn{1}{c}{3} & \multicolumn{1}{c}{5} & \multicolumn{1}{c}{10} 
  	& \multicolumn{1}{c}{25} \\ \midrule 
                        1\ \       & 0.90457                & 0.89391                & 0.91276                 & 0.90870 \\
                        3\ \       & 0.92696                & 0.93348                & 0.93630                 & 0.93587 \\
                        6\ \       & 0.93087                & 0.94000                & 0.94109                 & 0.94022 \\ 
\midrule\midrule
\end{tabular}
\end{table}

\subsection{Adversarial Results}\label{sect:adverse}

We trained a GAN to generate adversarial samples, and using these samples to
determine whether our deep learning model is robust under adversarial attack. 
First, we determine how close our GAN-generated data is to real data through
a simulated poisoning attack. Then we test how well the adversarial samples can evade 
an authenticator model.

As outlined in Section~\ref{chap:poison}, we poison our training dataset with 
an increasing percentage of fake data. All real samples are always included, 
so any changes in classification accuracy should be caused by our adversarial (fake) samples. 
We trained our DC-GAN for different numbers of epochs and different numbers of adversarial samples. 
Generally, a higher number of epochs results in better imitations of the original data. 
Increasing the number of adversarial samples in the training dataset gauges how well 
our 1D-CNN can resist large-scale poisoning.

The results of our poisoning attacks are given in Table~\ref{tab:adverse}. 
The accuracy remains relatively high at more than~90\%, 
even for high training epochs and up to a~1:1 ratio of real to fake data. 
Classification accuracy does decrease slightly when there are more 
adversarial samples in the training dataset, but the loss in accuracy is not large. 
These results show that our 1D-CNN is highly resistant to poisoning attacks.

\begin{table}[!htb]
\advance\tabcolsep by 2pt
\caption{Adversarial attack results (1840 real samples)}\label{tab:adverse}
\centering
\begin{tabular}{ccllll}
\midrule\midrule
\multirow{2}{*}{$\mbox{adversarial}_{\vphantom N}\atop\mbox{samples}^{\vphantom N}$} 
  & \multicolumn{4}{c}{epochs} \\ \cmidrule{2-5}
  & 10 & 25 & 50 & 100 \\ \midrule
  \z100                   & 0.93609                 & 0.93130                 & 0.94761                 & 0.94565 \\
  \z250                   & 0.94587                 & 0.94848                 & 0.93783                 & 0.94196 \\
  \z600                   & 0.94087                 & 0.93978                 & 0.94002                 & 0.94152 \\
 1840                  & 0.94000                 & 0.94152                 & 0.92239                 & 0.93435 \\ 
 \midrule\midrule
\end{tabular}
\end{table}

We conjecture that the reason for the limited success of our adversarial attack is that it is
exceedingly difficult to generate realistic signatures of the type considered in this paper.
While the real signatures in our dataset vary wildly, those that we generated using DC-GAN
appear to exhibit more homogeneity. We plan to investigate this issue further
in future work.

\section{Conclusion and Future Work}\label{chap:conclusion}

Previous research has shown that SVMs are a viable techniques for accelerometer-based gesture authentication~\cite{huang2021new}. In this paper, we expanded on and improved upon previous work. 
First, we refined the feature selection process with SVM-RFE to select the best features, 
while maintaining a high classification accuracy. Then, we used deep learning models, 
specifically 1D-CNNs, for classification. We obtained strong results, with greater 
than~90\%\ classification accuracy, slightly surpassing the accuracy of our SVM model. 
Lastly, we experimented with adversarial attacks on our 1D-CNN model, namely, 
poisoning and evasion attacks. These simulate realistic attacks, assuming an intruder has access 
to the real data, but not the model itself. 
Our results indicate that our 1D-CNN is robust under such attacks, 
achieving greater than~90\%\ accuracy for poisoning attacks and near perfect accuracy 
for evasion attacks.  

For future work, additional machine learning techniques could be considered. 
For example, long-short term memory (LSTM) models
could be used instead of 1D-CNNs, since LSTMs generally perform well on sequential data. 
Additionally, as mentioned in Section~\ref{sect:adverse}, 
we would like to explore alternative methods of generating adversarial 
samples to determine whether we can improve on the limited adversarial attack 
results obtained with DC-GANs.

\bibliographystyle{plain}
\bibliography{references.bib}

\end{document}

%% file: preamble.tex
\usepackage[scr=boondoxo,scrscaled=1.05]{mathalfa}
\usepackage[pdftex]{graphicx}
\usepackage{amsmath,amssymb,hhline,tikz,mathtools}
\usepackage{enumerate}
\usetikzlibrary{shapes.geometric,decorations.markings}
\usetikzlibrary{calc}
\usepackage[title]{appendix}
\usepackage{etoolbox}
\patchcmd{\appendices}{\quad}{: }{}{}
\usepackage{algorithm}
\usepackage{algorithmicx}
\usepackage{algpseudocode}
\usepackage{rotating}
\usepackage{afterpage}
\usepackage{xstring}
\usepackage{xcolor}
\usepackage{cmap}
\usepackage{mathdots}
\usepackage{colortbl}
\usepackage{booktabs}
\usepackage{textcomp}     
\usetikzlibrary{backgrounds}
\usepackage{pgfplots}
\pgfplotsset{compat=1.12}
\pgfmathsetseed{5112}
\usepackage{multirow}
\usepackage{pgfplotstable}
\usepackage{adjustbox}
\usepackage{diagbox}

\PassOptionsToPackage{hyphens}{url}
\usepackage{hyperref}
\hypersetup{
    pdftitle={Word Embedding Techniques for Malware Classification},
    pdfauthor={Mark Stamp},
    bookmarksnumbered=true,     
    bookmarksopen=true,         
    bookmarksopenlevel=3,       
    colorlinks=true,linkcolor=black,citecolor=black,urlcolor=blue,filecolor=blue,
    pdfstartview=Fit,           
    pdfpagemode=UseOutlines,
    pdfpagelayout=SinglePage
}

\definecolor{darkgreen}{rgb}{0.125,0.5,0.169}

\algrenewcommand{\algorithmiccomment}[1]{{\color{red}{\tt //}\ #1}}
\algnewcommand{\Initialize}[1]{%
  \State \textbf{Initialize:}
  \Statex \hspace*{\algorithmicindent}\parbox[t]{.8\linewidth}{\raggedright #1}
}
\algnewcommand{\Given}[1]{%
  \State \textbf{Given:}
  \Statex \hspace*{\algorithmicindent}\parbox[t]{.8\linewidth}{\raggedright #1}
}

\long\def\symbolfootnotetext[#1]#2{\begingroup%
\def\thefootnote{\fnsymbol{footnote}}\footnotetext[#1]{#2}\endgroup}

\let\oldsqrt\sqrt
\def\sqrt{\mathpalette\DHLhksqrt}
\def\DHLhksqrt#1#2{%
\setbox0=\hbox{$#1\oldsqrt{#2\,}$}\dimen0=\ht0
\advance\dimen0-0.2\ht0
\setbox2=\hbox{\vrule height\ht0 depth -\dimen0}%
{\box0\lower0.4pt\box2}}

\def\clap#1{\hbox to 0pt{\hss#1\hss}}

\allowdisplaybreaks

\def\figureFastSpeed{s}\def\figureSpeed{f}
\let\figureFastSpeed=\figureSpeed
\def\selectFigureSpeed#1#2{
\if\figureSpeed\figureFastSpeed #1\else #2\fi}

\def\srowvecc#1#2{(\!\begin{array}{cc} 
      \noexpandarg\IfBeginWith{#1}{-}{\! #1}{#1}
    & #2\kern-0.5pt\end{array}\!)}
\def\rowvecc#1#2{\left(\!\begin{array}{cc} 
      \noexpandarg\IfBeginWith{#1}{-}{\! #1}{#1}
    & #2\kern-0.5pt\end{array}\!\right)}
\def\rowveccc#1#2#3{\left(\!\begin{array}{ccc} 
      \noexpandarg\IfBeginWith{#1}{-}{\! #1}{#1}
    & #2 
    & #3\kern-0.5pt\end{array}\!\right)}
\def\rowvecccc#1#2#3#4{\left(\!\begin{array}{cccc}
      \noexpandarg\IfBeginWith{#1}{-}{\! #1}{#1}
    & #2 
    & #3 
    & #4\kern-0.5pt\end{array}\!\right)}
\def\srowvecccc#1#2#3#4{\bigl(\!\begin{array}{cccc}
      \noexpandarg\IfBeginWith{#1}{-}{\! #1}{#1}
    & #2 
    & #3 
    & #4\kern-0.5pt\end{array}\!\bigr)}
\def\rowveccccc#1#2#3#4#5{\left(\!\begin{array}{ccccc} 
      \noexpandarg\IfBeginWith{#1}{-}{\! #1}{#1}
    & #2
    & #3
    & #4
    & #5\kern-0.5pt\end{array}\!\right)}
\def\srowvecccccc#1#2#3#4#5#6{(\!\begin{array}{cccccc} 
      \noexpandarg\IfBeginWith{#1}{-}{\! #1}{#1}
    & #2
    & #3
    & #4
    & #5
    & #6\kern-0.5pt\end{array}\!)}
\def\rowvecccccc#1#2#3#4#5#6{\left(\!\begin{array}{cccccc} 
      \noexpandarg\IfBeginWith{#1}{-}{\! #1}{#1}
    & #2
    & #3
    & #4
    & #5
    & #6\kern-0.5pt\end{array}\!\right)}

\def\figureType{*}\def\figureSlowType{slowType}
\def\selectFigureType#1#2{
\if\figureType\figureSlowType #1\else #2\fi}
\makeatletter
\newcommand{\lowsub}[1]{\mathpalette{\raisem@th{#1}}}
\newcommand{\raisem@th}[3]{\raisebox{-#1}{$#2#3$}}
\makeatother
\def\halfthin{\kern 0.083em}

\pgfkeys{
    /pgf/number format/precision=2, 
    /pgf/number format/fixed zerofill=true }
    
\pgfplotstableset{
    /color cells/min/.initial=0,
    /color cells/max/.initial=1000,
    /color cells/textcolor/.initial=,
    color cells/.code={%
        \pgfqkeys{/color cells}{#1}%
        \pgfkeysalso{%
            postproc cell content/.code={%
                \begingroup
                \pgfkeysgetvalue{/pgfplots/table/@preprocessed cell content}\value
\ifx\value\empty
\endgroup
\else
                \pgfmathfloatparsenumber{\value}%
                \pgfmathfloattofixed{\pgfmathresult}%
                \let\value=\pgfmathresult
                \pgfplotscolormapaccess
                    [\pgfkeysvalueof{/color cells/min}:\pgfkeysvalueof{/color cells/max}]%
                    {\value}%
                    {\pgfkeysvalueof{/pgfplots/colormap name}}%
                \pgfkeysgetvalue{/pgfplots/table/@cell content}\typesetvalue
                \pgfkeysgetvalue{/color cells/textcolor}\textcolorvalue
                \toks0=\expandafter{\typesetvalue}%
                \xdef\temp{%
                    \noexpand\pgfkeysalso{%
                        @cell content={%
                            \noexpand\cellcolor[rgb]{\pgfmathresult}%
                            \noexpand\definecolor{mapped color}{rgb}{\pgfmathresult}%
                            \ifx\textcolorvalue\empty
                            \else
                                \noexpand\color{\textcolorvalue}%
                            \fi
                            \the\toks0 %
                        }%
                    }%
                }%
                \endgroup
                \temp
\fi
            }%
        }%
    }
}

\makeatletter
\newcommand*\bigcdot{\mathpalette\bigcdot@{.5}}
\newcommand*\bigcdot@[2]{\mathbin{\vcenter{\hbox{\scalebox{#2}{$\m@th#1\bullet$}}}}}
\makeatother

\def\k{\kern 2.75pt}

\newlength{\xxxxx}
\def\z{\phantom{0}}

%% file: figures/kernel_trick.tex
  \begin{tabular}{ccc}
  \begin{tikzpicture} 
    \begin{axis}[width=0.5\textwidth,height=0.5\textwidth,xmin=-3,xmax=3,ymin=-3,ymax=3] 
      \pgfplotsset{ticks=none}      
       \addplot[color=red, mark=*,only marks,mark size=1.5] coordinates { 
(0.331578, 0.746603)
(-0.394443, -0.263352)
(0.625403, -0.375284)
(0.911628, -0.359273)
(0.500405, 0.703271)
(-0.231736, 0.872177)
(0.170166, -0.565556)
(-0.274232, -0.834498)
(-0.334136, -0.078180)
(-0.592072, 0.089388)
(-0.075179, -0.030616)
(0.041852, -0.825840)
(-0.732711, -0.394369)
(0.511456, -0.000657)
(0.832060, 0.272873)
      };
      \addplot[color=blue, mark=square,fill=blue,only marks,mark size=1.5] coordinates { 
(2.390971, 0.052211)
(-1.437509, -1.752351)
(1.886234, -1.604676)
(2.395163, 0.054876)
(2.288330, -0.765636)
(0.807904, 1.929355)
(0.207518, 2.449499)
(1.900863, -0.937399)
(-0.548196, -1.621026)
(-1.322488, 2.083358)
(1.006897, 2.228109)
(1.735974, 0.874161)
(1.838050, -1.491165)
(0.600720, 2.362412)
(1.456580, -1.733536)
(-0.316657, 2.221768)
(2.087640, 1.203409)
(1.616520, 1.387743)
(-2.433935, -0.209267)
(-1.788833, -0.700340)
(-0.877470, -1.504627)
(-1.107672, -1.537246)
(1.184006, 2.095965)
(-0.080668, -2.371692)
(2.414325, 0.371814)
(0.554228, 2.404566)
(1.505664, 1.283217)
(-1.162894, -1.890363)
(-2.098915, -0.022764)
(1.920753, 1.102422)
(-2.288927, -0.565147)
(1.905982, -1.506967)
(2.233564, -0.516066)
(-2.003422, 0.863837)
(-1.910277, -0.269332)
(0.966902, -2.244612)
(-0.568292, 1.923142)
(0.112558, 2.219818)
(-1.549785, 0.692908)
(-2.193879, 0.538788)
(-1.818708, 1.326454)
(2.270850, -0.798137)
(-2.092595, 1.043138)
(2.155481, -0.168913)
(-0.017986, -2.319378)
(1.239888, 1.707677)
(-0.473226, -1.885332)
(0.537416, 1.709329)
(-1.739303, -1.065284)
(0.657574, -1.519209)
      };
    \end{axis}
  \end{tikzpicture}  
  & \raisebox{0.475in}{${{\displaystyle\phi} \atop {\displaystyle\implies}}$} &
    \begin{tikzpicture}
    \begin{axis}[width=0.5\textwidth,height=0.5\textwidth,xmin=-3,xmax=3,ymin=-3,ymax=3,zmin=0,zmax=5] 
      \pgfplotsset{ticks=none}
      \addplot3[color=red, mark=*,only marks,mark size=1.5] coordinates { 
(0.331578, 0.746603,0.6)
(-0.394443, -0.263352,0.6)
(0.625403, -0.375284,0.6)
(0.911628, -0.359273,0.6)
(0.500405, 0.703271,0.6)
(-0.231736, 0.872177,0.6)
(0.170166, -0.565556,0.6)
(-0.274232, -0.834498,0.6)
(-0.334136, -0.078180,0.6)
(-0.592072, 0.089388,0.6)
(-0.075179, -0.030616,0.6)
(0.041852, -0.825840,0.6)
(-0.732711, -0.394369,0.6)
(0.511456, -0.000657,0.6)
(0.832060, 0.272873,0.6)
      };
      \addplot3[color=blue, mark=square,fill=blue,only marks,mark size=1.5] coordinates { 
(2.390971, 0.052211,4.0)
(-1.437509, -1.752351,4.0)
(1.886234, -1.604676,4.0)
(2.395163, 0.054876,4.0)
(2.288330, -0.765636,4.0)
(0.807904, 1.929355,4.0)
(0.207518, 2.449499,4.0)
(1.900863, -0.937399,4.0)
(-0.548196, -1.621026,4.0)
(-1.322488, 2.083358,4.0)
(1.006897, 2.228109,4.0)
(1.735974, 0.874161,4.0)
(1.838050, -1.491165,4.0)
(0.600720, 2.362412,4.0)
(1.456580, -1.733536,4.0)
(-0.316657, 2.221768,4.0)
(2.087640, 1.203409,4.0)
(1.616520, 1.387743,4.0)
(-2.433935, -0.209267,4.0)
(-1.788833, -0.700340,4.0)
(-0.877470, -1.504627,4.0)
(-1.107672, -1.537246,4.0)
(1.184006, 2.095965,4.0)
(-0.080668, -2.371692,4.0)
(2.414325, 0.371814,4.0)
(0.554228, 2.404566,4.0)
(1.505664, 1.283217,4.0)
(-1.162894, -1.890363,4.0)
(-2.098915, -0.022764,4.0)
(1.920753, 1.102422,4.0)
(-2.288927, -0.565147,4.0)
(1.905982, -1.506967,4.0)
(2.233564, -0.516066,4.0)
(-2.003422, 0.863837,4.0)
(-1.910277, -0.269332,4.0)
(0.966902, -2.244612,4.0)
(-0.568292, 1.923142,4.0)
(0.112558, 2.219818,4.0)
(-1.549785, 0.692908,4.0)
(-2.193879, 0.538788,4.0)
(-1.818708, 1.326454,4.0)
(2.270850, -0.798137,4.0)
(-2.092595, 1.043138,4.0)
(2.155481, -0.168913,4.0)
(-0.017986, -2.319378,4.0)
(1.239888, 1.707677,4.0)
(-0.473226, -1.885332,4.0)
(0.537416, 1.709329,4.0)
(-1.739303, -1.065284,4.0)
(0.657574, -1.519209,4.0)
      };
    \end{axis}
  \end{tikzpicture}  
  \end{tabular}

%% file: figures/barMulticlassAccuracy.tex
\begin{tikzpicture}[scale=0.95, every node/.style={scale=1.0}]
\begin{axis}[
        width  = 0.85*\textwidth,
        height = 7.5cm,
        ymin=0.0,ymax=1.10,
        ytick={0.0,0.20,0.40,0.60,0.80,1.00},
        major x tick style = transparent,
        ybar=5*\pgflinewidth,
        bar width=10.0pt,
        xlabel = {Features},
        ylabel = {Accuracy},
        symbolic x coords={1, 2, 3, 4, 5},
        xticklabels={{$L,\mu,m,\sigma,k,s$},{$\mu,m,\sigma,k,s$},{$\mu,m,\sigma,s$},
        		{$\mu,m,\sigma(y,z),s(y)$},{$\mu(x,z),m,s(y)$}},
	y tick label style={
    		/pgf/number format/.cd,
   		fixed,
   		fixed zerofill,
    		precision=2},
        xtick = data,
        x tick label style={
        		rotate=60,
		font=\small,
		anchor=north east,
		inner sep=0mm
		},
        nodes near coords,
        every node near coord/.append style={rotate=90, 
        								   anchor=west,
								   font=\scriptsize,
								   /pgf/number format/.cd,
								   fixed,
								   fixed zerofill,
								   precision=2},
        enlarge x limits=0.15,
      bar shift={-1.125*\pgfplotbarwidth}
]
\addplot [fill=blue,opacity=1.00]
coordinates {
(1, 0.95)
(2, 0.94)
(3, 0.92)
(4, 0.89)
(5, 0.76)
};
\end{axis}
\begin{axis}[
        width  = 0.85*\textwidth,
        height = 7.5cm,
        ymin=0.0,ymax=0.27,
        ytick={0.0,0.05,0.10,0.15,0.20,0.25},
        major x tick style = transparent,
        ybar=5*\pgflinewidth,
        bar width=10.0pt,
        axis x line=none,
        axis y line*=right,
        ylabel = {FAR and FRR},
        symbolic x coords={1, 2, 3, 4, 5},
        xticklabels={{$L,\mu,m,\sigma,k,s$},{$\mu,m,\sigma,k,s$},{$\mu,m,\sigma,s$},
        		{$\mu,m,\sigma(y,z),s(y)$},{$\mu(x,z),m,s(y)$}},
	y tick label style={
    		/pgf/number format/.cd,
   		fixed,
   		fixed zerofill,
    		precision=2},
        xtick = data,
        x tick label style={
        		rotate=60,
		font=\small,
		anchor=north east,
		inner sep=0mm
		},
        nodes near coords,
        every node near coord/.append style={rotate=90, 
        								   anchor=west,
								   font=\scriptsize,
								   /pgf/number format/.cd,
								   fixed,
								   fixed zerofill,
								   precision=4},
        enlarge x limits=0.15,
]
\addplot [fill=red,opacity=1.00]
coordinates {
(1, 0.0014)
(2, 0.0013)
(3, 0.0018)
(4, 0.0024)
(5, 0.0053)
};
\end{axis}
\begin{axis}[
        width  = 0.85*\textwidth,
        height = 7.5cm,
        ymin=0.0,ymax=0.27,
        ytick={0.0,0.05,0.10,0.15,0.20,0.25},
        major x tick style = transparent,
        ybar=5*\pgflinewidth,
        bar width=10.0pt,
        axis x line=none,
        axis y line=none,
        ylabel = {FAR and FRR},
        symbolic x coords={1, 2, 3, 4, 5},
        xticklabels={{$L,\mu,m,\sigma,k,s$},{$\mu,m,\sigma,k,s$},{$\mu,m,\sigma,s$},
        		{$\mu,m,\sigma(y,z),s(y)$},{$\mu(x,z),m,s(y)$}},
	y tick label style={
    		/pgf/number format/.cd,
   		fixed,
   		fixed zerofill,
    		precision=2},
        xtick = data,
        x tick label style={
        		rotate=60,
		font=\small,
		anchor=north east,
		inner sep=0mm
		},
        nodes near coords,
        every node near coord/.append style={rotate=90, 
        								   anchor=west,
								   font=\scriptsize,
								   /pgf/number format/.cd,
								   fixed,
								   fixed zerofill,
								   precision=3},
        enlarge x limits=0.15,
        legend cell align=left,
        legend columns=-1,
        legend style={
                at={(0.725,1.025)},
                anchor=south,
                column sep=1ex
        },
      bar shift={1.125*\pgfplotbarwidth}
]
\addlegendimage{no markers,black,fill=blue}
\addlegendentry{Accuracy}
\addlegendimage{no markers,black,fill=red}
\addlegendentry{FAR}
\addplot [fill=yellow,opacity=1.00]
coordinates {
(1, 0.057)
(2, 0.059)
(3, 0.083)
(4, 0.11)
(5, 0.22)
};
\addlegendentry{FRR}
\end{axis}
\end{tikzpicture}